# Interaction-Free Measurements: A Complex Nonlinear Explanation

A. Cardoso[1], J. L. Cordovil[1] and J. R. Croca[1,2]

1 – Centro de Filosofia das Ciências da Universidade de Lisboa, Faculdade de Ciências da Universidade de Lisboa, Campo Grande, Edifício C4, 1749-016 Lisboa, Portugal
2 – Departamento de Física da Faculdade de Ciências da Universidade de Lisboa, Campo Grande, Edifício C8, 1749-016 Lisboa, Portugal

Abstract: In this paper we show that interaction-free measurements, which have been object of much discussion in the last few decades, can be explained in a natural and intuitive way in the framework of complex nonlinear quantum physics, contrary to what is claimed by some orthodox authors that present them as incomprehensible, mind-boggling experiments.

Keywords: nonlinear quantum physics, orthodox quantum mechanics, interaction-free measurements

## 1. Introduction

The so-called interaction-free measurements have been object of much discussion[1] in the last few decades and have been presented in the literature as mysterious, almost supernatural phenomena. Like illusionists on stage showing off their tricks to an audience, most authors present their results in such an elaborated way that it becomes impossible for the reader to understand what is actually happening in the experiments. However, as with every magic trick, all these experiments can be explained in a very simple manner if we are allowed to look behind the curtain. In the core of this issue is of course the inability of orthodox quantum mechanics to provide us with a process through which these measurements are actually made. However, as we will show in this paper, interaction-free measurements can be easily explained if we are willing to take a complex nonlinear approach.

Let us start our discussion by looking at one of the earliest thought experiments of this kind, proposed by Renninger[2] more than half a century ago. All the other interaction-free measurement experiments presented in the literature may be seen as particular and more complicated versions of this experiment. The so-called Renninger negative-result experiment can be presented using the arrangement shown in Fig. 1.

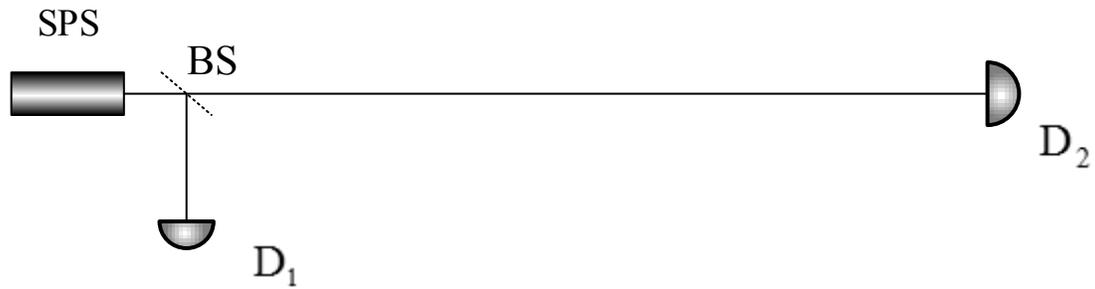

Fig. 1 – Renninger negative-result experiment.

When a photon, whose state can be represent by a wave-function $\psi$, emitted by a single photon source SPS reaches the 50% beam-splitter BS, it will be split into two waves $\psi_1$ and $\psi_2$, each one with half the amplitude of the original wave such that $\psi = \psi_1 + \psi_2$. One of them, $\psi_1$, will be reflected and will take the (shorter) route to reach detector $D_1$ after a time interval $\Delta t_1$ and the other, $\psi_2$, will be transmitted and will take the (longer) route to arrive at detector $D_2$ after an interval $\Delta t_2$. As a consequence, 50% of the photons will be detected by $D_1$ and the other 50% will reach $D_2$ (for the purposes of the thought experiment the detectors are assumed to be 100% efficient). In this case, it is clear that if a detection is made by $D_1$ then we know for sure that no particle will reach $D_2$. Conversely, if no detection is made at $D_1$ after a time interval $\Delta t_1$ then we know that a photon will reach detector $D_2$ after the larger time interval $\Delta t_2$.

Now, according to the orthodox interpretation of quantum mechanics[3], before any measurement is made the system is in a linear superposition of two states whose probabilities are represented by the waves $\psi_1$ and $\psi_2$. In these circumstances two cases may arise:

1 – A 'positive' detection is made at $D_1$. In this case the system will collapse into the state represented by $\psi = \psi_1$ and consequently $\psi_2 = 0$, therefore no photon will reach $D_2$.

2 – No detection is made at $D_1$ after a time $\Delta t_1$. According to this interpretation the system will collapse into the state $\psi = \psi_2$. It is worth noting that this collapse occurs before the photon is actually seen at detector $D_2$, i.e., there is a collapse of the wave function without any detection. This is therefore called a 'negative' measurement.

The assumption that the photon is not in a well-defined state until a measurement is made implies, in the second case, that this measurement will 'magically' make the system collapse into one of the possible states even though no physical interaction occurred.

The situation looks even stranger if we slightly modify the setup by connecting a light to detector $D_1$ and adding a long fiber-optical cable that will dramatically increase the length of the path to detector $D_2$, in front of which we then place a screen detector (see Fig. 2).

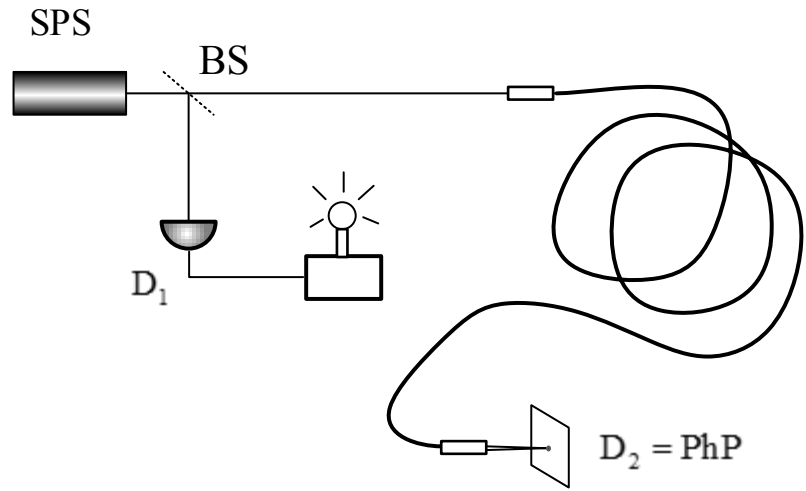

Fig. 2 – Modified Renninger negative-result experiment.

As we have just discussed, if a photon reaches $D_1$ after a time interval $\Delta t_1$ then the detector will be triggered, and the light will be on. If, however, the light stays off after a time $\Delta t_1$ then we can conclude that the photon took the much longer path to detector $D_2$ and thus a dot will eventually appear on the detector screen.

A question then arises: since in this case, according to the orthodox interpretation, the system collapses into the state $\psi_2$ just after the time interval $\Delta t_1$ then will a dot appear on the detector screen at that moment? The answer will clearly be ''no'', as nothing has yet been detected at $D_2$. Nevertheless, even if no physical interaction occurred we know for sure that at a later time $\Delta t_2$ the photon shall be revealed at the distant detector $D_2$. Even though it has been established that the probability of $D_2$ making a detection is 1, somehow we may have to way for a very long time before the probability wave actually arrives at the detector.

This strange problem, raised by the orthodox interpretation of quantum mechanics, can be easily solved in the framework of nonlinear quantum physics[4] inspired in the early ideas proposed by de Broglie[5]. In this approach to understand Nature, the photon, or any quantum particle, is composed of an extended real wave – the guiding wave, theta wave or subquantum wave – plus a highly energetic and very well localized kernel, corpuscle or acron. This real physical wave guides the corpuscle according to the principle of eurhythmy, i.e., preferentially to the regions were the intensity of the wave is greater.

According to this complex nonlinear approach, the $\psi$ wave describing the particle will be split into two as it arrives at the beamsplitter BS. The indivisible kernel or acron, however, will take only one of the paths with equal probability. Therefore, one of the waves $\psi_1$ and $\psi_2$ will carry no acron.

Thus, if a detection is made at $D_1$ then it means that the wave transmitted through the longer path is travelling alone, without the corpuscle, and therefore will not have enough energy to trigger detector $D_2$. Conversely, if no detection is made at $D_1$ after a time $\Delta t_1$ then we know that the wave taking the longer path is carrying the corpuscle, which will trigger detector $D_2$ after the time interval $\Delta t_2$. In this case there is clearly no mystery involved in the measurement of a particle.

In the following we will look at two other interaction-free measurement experiments proposed in the literature and discuss them in the framework of orthodox quantum mechanics as well as in complex nonlinear terms. In both experiments we will show that the measurement can be explained through an actual, physical interaction if we take the second interpretation.

## 2. The Elitzur-Vaidman bomb tester

The thought experiment known as the Elitzur-Vaidman bomb tester[6], proposed by Elitzur and Vaidman in 1993 and tested by Kwiat, Weinfurter, Herzog, Zeilinger and Kasevich two years later, employs a Mach-Zehnder interferometer, a light source SPS that emits one photon at a time and a bomb (or an opaque object) that is placed in the lower arm of the apparatus. Let us first discuss the basic interferometer, with no bomb, whose scheme is shown in Fig. 3.

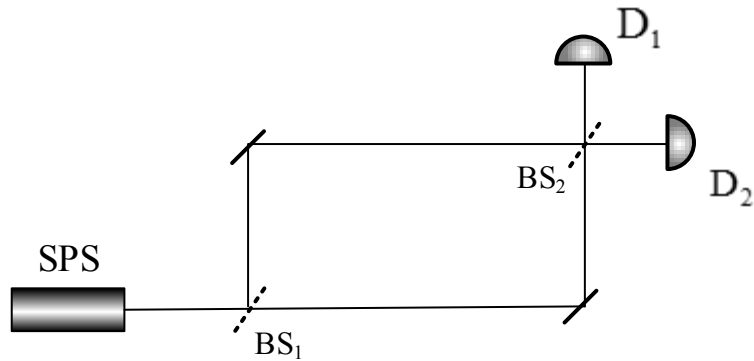

Fig. 3 – Mach-Zehnder interferometer.

As in the previous experiment, when a single emitted photon reaches the first 50% beam-splitter, $BS_1$, its wave $\psi$ will be split into two waves $\psi_1$ and $\psi_2$ with half the amplitude of the original wave. One of them, $\psi_1$, will be reflected (taking the upper route) and the other, $\psi_2$, will be transmitted (taking the lower route). The same will happen to both waves as they simultaneously reach – assuming that the upper and lower arms of the apparatus have equal optical paths – the second beamsplitter, $BS_2$, each one of them being again divided into two waves (one transmitted and the other reflected) with a quarter of the amplitude of the original wave. Therefore, both detectors will receive half of the upper and lower waves, which will then interfere.

The waves that reach detector $D_2$ will interfere constructively and the ones reaching detector $D_1$ will interfere destructively, which means that all photons will be detected at $D_2$

and none at $D_1$. In this situation the expected intensities seen at each detector will be

$$\begin{cases} I_1 = 0 \\ I_2 = I_0 \end{cases} \qquad (2.1)$$

where $I_1$ and $I_2$ are the intensities seen by the detectors and $I_0$ the intensity of the source for a lossless interferometer.

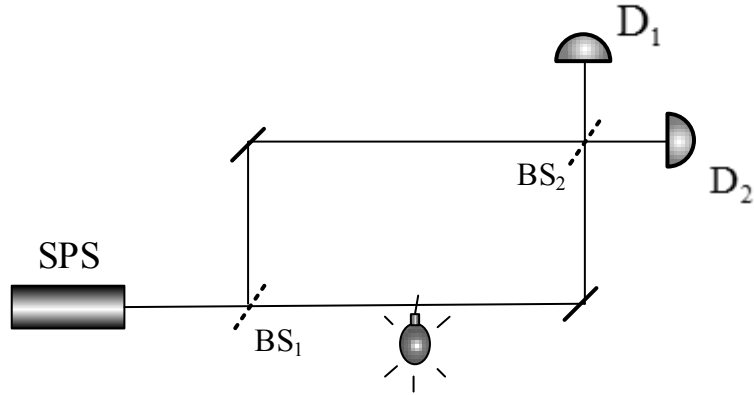

Fig. 4 – Elitzur-Vaidman bomb tester.

## 2.1. Orthodox interpretation

We will now introduce the bomb in the interferometer and describe the Elitzur-Vaidman experiment from the orthodox point of view, as it is usually done in the literature. If the bomb is blocking the lower path, as shown in Fig. 4, the possible results will be the following:

1 – The bomb is fake (i.e., it does not have a fuse). In such situation, everything happens as if there was no bomb. Consequently, since both paths are open and thus $\psi = \psi_1 + \psi_2$, no changes are expected and detector $D_2$ will detect all the photons.

2 – The bomb is usable. In this situation two cases may occur:

a) The photon takes the lower route and interacts with the bomb, which will then explode, making $\psi = \psi_2$.

b) The photon takes the upper path. In this case, since the bomb is blocking the lower path, the wave function will collapse and the only probability wave remaining will be the one taking the upper route, $\psi = \psi_1$. Consequently there will be no interference and the photon will have equal probabilities of reaching detectors $D_1$ and $D_2$. In this case the expected intensities observed will be

$$I_1 = I_2 = \frac{1}{2} I_0 \qquad (2.2)$$

Thus, if a photon reaches detector $D_1$ then the bomb must be usable, and we know it even though no photon interacted with it. In sum, we have been able to know whether an object is present without interacting with it, and the only way to explain this result is by invoking Niels Bohr's complementary principle: if one of the two paths is not available, which tells us which path the particle took, then the interference pattern has to disappear.

The major problem with this interpretation is that it does not give us a physical mechanism through which the photon 'senses' the usable bomb. A probability wave, not being real, would not able to interact with the bomb and therefore the photon would not be able to know whether the bomb is blocking the lower path. As a consequence, from this point of view the experiment remains a mystery.

## 2.2. Complex nonlinear interpretation

The issue raised once more by the orthodox interpretation can again be easily solved if we look at the photon as a complex system composed of a real wave that guides the corpuscle through its path to the detector. In this case, the real guiding wave accompanying the photonic acron splits into two physical waves as it arrives at the beam-splitter $BS_1$. The corpuscle or high-energetic part, however, takes only one of the paths with equal probabilities. Therefore one of the real theta waves $\psi_1$ and $\psi_2$ will carry no acron and will travel alone throughout one of the paths.

If there is no bomb, or if the bomb is fake, then the waves will meet at the second beamsplitter $BS_2$ to interfere constructively at detector $D_2$ and destructively at detector $D_1$. The photon's high-energetic part will then take the path where the intensity of the superimposed waves is highest and will thus reach detector $D_2$.

If a real bomb is blocking the lower path then we have again two possible outcomes:

a) The corpuscle takes the lower path and interacts with the bomb, which shall then explode.

b) The corpuscle takes the upper path. In this case, the real subquantum wave taking the lower path does not carry enough energy to trigger the bomb, and it will either be absorbed or have its phase randomly changed, losing coherence with the wave taking the upper path. In either case there will be no interference at the second beam-splitter $BS_2$ and the photon will have equal probabilities of reaching detectors $D_1$ and $D_2$.

As we see from the previous discussion, the existence of a real wave allows us to explain the process through which the corpuscle 'senses' the bomb without directly interacting with it. The answer is straightforward: even if the empty real wave does interact with the bomb either by being absorbed or phase-randomized, it has no direct action on the overall behavior of the acron following the upper path. It is therefore possible in this way, contrary to what is claimed in most of the literature, to look behind the curtain and solve the mysteries created by the orthodox interpretation.

## 3. Mandel-Zeilinger interferometer

We will now look at another, similar experiment done in 1991 by Cow, Wang and Mandel and more recently repeated by Lemos, Borish, Cole, Ramelow, Lapkiewicz and Zeilinger[7]. In this experiment (shown in Fig. 5), in each of the two paths resulting from the first beam-splitter $BS_1$, the authors introduce a nonlinear crystal, which from an incident photon produces a pair of photons – one called signal and the other one idler – that are not coherent but are correlated in both space and time. The idler photon produced at the nonlinear crystal $NL_1$ is injected in the nonlinear crystal $NL_2$ together with the UV pump beam, inducing coherence between the two signal photons. The path of the idler photon created at the upper crystal $NL_1$ meets the path of the one created at the lower crystal $NL_2$ in a way that the two paths become coincident.

In the path between the two nonlinear crystals (i.e., the path of the idler photon created at $NL_1$) there is an object O that we can chose to be present or absent, which is somehow analogous to the bomb in the previous experiment.

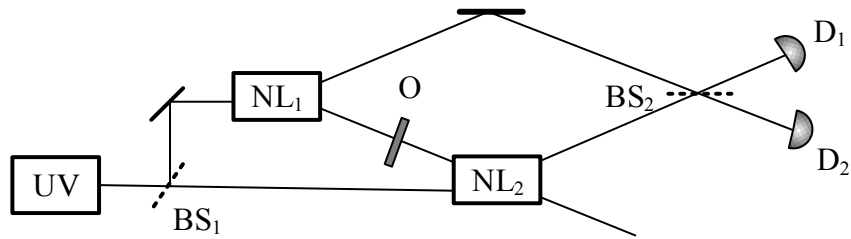

Fig. 5 – Mandel-Zeilinger interferometer.

## 3.1. Orthodox interpretation

According to orthodox quantum mechanics, and as it is viewed by the authors of the experiment, if no object O is present between the non-linear crystals then, as the paths of the two idler photons are coincident, there is no way to know whether a signal photon arriving at the second beamsplitter $BS_2$ came through the upper or the lower path, and thus the corresponding probability waves will interfere constructively as they arrive at detector $D_2$ and destructively at detector $D_1$, as in the previous experiment. Therefore, all the photons will be detected by $D_2$.

Now, if an opaque object O is introduced in the path between the two crystals then any idler photon created at the upper crystal $NL_1$ will be blocked, and so a detection of an idler photon (which we could do if we chose to) would mean that the signal photon detected in coincidence by $D_1$ or $D_2$ could only have been produced by the lower crystal $NL_2$, and therefore it must have arrived at the second beamsplitter through the lower path.

Conversely, if no idler photon was observed then the signal photon detected in coincidence could only have been produced at the upper non-linear crystal $NL_1$, meaning that it must have taken the upper path. This which-path information will thus destroy the interference and the signal photon will have equal probabilities of reaching detectors $D_1$ and $D_2$.

Once more, this interpretation does not explain the process through which the signal photon knows whether there is an object blocking the path between the two crystals, as there is no way they could have interacted. As in the bomb-testing experiment, we are led to conclude that the measurement was mysteriously done without any kind of interaction.

### 3.2. Complex nonlinear interpretation

If we take our nonlinear approach, the interpretation of this experiment is much more intuitive.

If there is no object O present in the setup then an idler photon produced at the nonlinear crystal $NL_1$ is able to reach the crystal $NL_2$, and this will induce a coherence between the signal photons produced at both crystals as well as between the corresponding idler photons (through a process called phase-locking). Therefore, a constructive interference will happen at detector $D_2$ (which will detected all the photons) and a destructive interference at detector $D_1$.

If there is an object in the path between the crystals then an idler photon leaving the nonlinear crystal $NL_1$ will not reach the lower crystal $NL_2$ and thus the signal photons will remain incoherent, which means that there will be no interference when they arrive at the detectors. Thus, again, any signal photon will have equal probabilities of reaching detectors $D_1$ and $D_2$.

As we see once more from this discussion, our complex nonlinear interpretation is able to explain the physical process through which the measurement of the object is made, without resorting to any kind of 'magic'.

### 4. Conclusion

In this paper we have shown, by looking at three different experiments, that interaction-free measurements can be easily explained in an intuitive way if we are willing to take a complex nonlinear approach. In all the three cases, this interpretation provides us with a physical process through which these measurements are actually made, allowing the reader to understand what is actually happening in the experiments and avoiding the mind-boggling explanations given by the orthodox interpretation.